\documentclass[runningheads]{llncs}

\usepackage{tabularx}
\usepackage{graphicx}
\usepackage{graphics}
\usepackage{booktabs}
\usepackage{amsmath}
\usepackage{fixltx2e}
\usepackage{subcaption}
\usepackage{wrapfig, blindtext}
\usepackage[dvipsnames]{xcolor,colortbl}
\usepackage{fontawesome}
\usepackage{chngcntr}
\usepackage{hyperref}

\definecolor{LightCyan}{rgb}{0.6,0.6,0.7}
\definecolor{DarkGray}{rgb}{0.75,0.75,0.8}
\definecolor{Gray}{rgb}{0.8,0.8,0.825}
\definecolor{LightGray}{rgb}{0.875,0.875,0.9}

\newcolumntype{g}{>{\columncolor{Gray}}r}
\newcolumntype{y}{>{\columncolor{LightGray}}c}
\newcolumntype{z}{>{\columncolor{DarkGray}}c}
\newcolumntype{h}{>{\columncolor{LightCyan}}r}

\usepackage{tikz}
\usepackage{collcell}
 
\newcommand*{\minVal}{0}%
\newcommand*{\maxVal}{1.0}%
 
\newcommand{\gradient}[1]{%
        \pgfmathsetmacro{\PercentColor}{100.0*(#1-\minVal)/(\maxVal-\minVal)}
        \hspace{-0.33em}\colorbox{RoyalBlue!\PercentColor!White}{#1}
}

\newcolumntype{G}{>{\collectcell\gradient}c<{\endcollectcell}}

\newcommand{\sectionnotitle}[1]{%
  \par\refstepcounter{section}
  \sectionmark{#1}
  \addcontentsline{toc}{section}{\protect\numberline{\thesection}#1}
}

\begin{document}

\title{Validating Simulations of User Query Variants}

\author{Timo Breuer\inst{1} \and
        Norbert Fuhr\inst{2} \and
        Philipp Schaer\inst{1}}

\authorrunning{Breuer et al.}

\institute{TH Köln, Germany \\
\email{firstname.lastname@th-koeln.de} \and
Universität Duisburg-Essen, Germany \\
\email{norbert.fuhr@uni-due.de}}

\maketitle              

\begin{abstract}
System-oriented IR evaluations are limited to rather abstract understandings of real user behavior. As a solution, simulating user interactions provides a cost-efficient way to support system-oriented experiments with more realistic directives when no interaction logs are available. While there are several user models for simulated clicks or result list interactions, very few attempts have been made towards query simulations, and it has not been investigated if these can reproduce properties of real queries. In this work, we validate simulated user query variants with the help of TREC test collections in reference to real user queries that were made for the corresponding topics. Besides, we introduce a simple yet effective method that gives better reproductions of real queries than the established methods. Our evaluation framework validates the simulations regarding the retrieval performance, reproducibility of topic score distributions, shared task utility, effort and effect, and query term similarity when compared with real user query variants. While the retrieval effectiveness and statistical properties of the topic score distributions as well as economic aspects are close to that of real queries, it is still challenging to simulate exact term matches and later query reformulations.
\end{abstract}

\section{Introduction}
In accordance with the Cranfield paradigm, the underlying user model of system-oriented IR evaluations is an abstract representation of real search behavior. The simplified understanding of users is limited to a single query and the examination of the result list in its entirety. Real search behavior is more complex: searching is normally an iterative process with query reformulations, and not every search result is examined but rather picked out after judging its snippet text. To compensate for this shortcoming, it is common practice to include (logged) user interactions in the evaluation process. Industrial research is often supported by large datasets of user interactions that, unfortunately, cannot be shared publicly, e.g., due to privacy concerns \cite{DBLP:conf/cikm/CraswellCMYB20}. Carterette et al. address the lack of user interaction data available to academic research by introducing the concept of Dynamic Test Collections \cite{DBLP:conf/ictir/CarteretteBZ15}. Their framework expands test collections with simulated interactions comprising the entire sequence of interactions, including the simulation of queries, clicks, dwell times, and session abandonment.

Our work can be seen in the light of Dynamic Test Collections, but with a special focus on simulating user query variants (UQVs). While previous work on simulating interactions either focused on the completeness of interaction sequences \cite{DBLP:conf/ictir/CarteretteBZ15,DBLP:conf/cikm/MaxwellA16,DBLP:conf/ictir/ZhangLZ17}, click interactions \cite{DBLP:series/synthesis/2015Chuklin}, or stopping rules \cite{DBLP:conf/cikm/MaxwellA16,DBLP:journals/ir/PaakkonenKKAMJ17}, work on simulating queries is underrepresented \cite{gunther2021assessing}. To the best of our knowledge, the degree to which query simulators reproduce real user queries has not yet been analyzed with TREC test collections. As opposed to previous work in this regard, it is not our primary goal to generate the most effective queries but rather to validate simulated queries, since the query formulation is one of the first user interactions with the search system and as such it is a critical component for any subsequent simulated interactions like clicks and others. More specifically, our evaluations answer \textbf{(RQ1)} \textit{How do real user queries relate to simulated queries made from topic texts and known-items in terms of retrieval effectiveness?} and \textbf{(RQ2)} \textit{To which degree do simulated queries reproduce real queries provided that only resources of the test collection are considered for the query simulation?}

Our contributions are as follows. \textbf{(1)} We introduce an evaluation framework that is used to analyze to which extent simulations reproduce real queries and that reveals current limitations, \textbf{(2)} we compare and analyze conventional query simulation methods that do not rely on large-scale interaction logs, \textbf{(3)} we propose a new simulation method and hereby show that the parameterized query reformulation behavior results in a better approximation of real queries and resembles those of specific users, \textbf{(4)} we publish the code of the experiments and provide the simulated queries for follow-up studies.

\section{Related Work}
Carterette et al. introduced Dynamic Test Collections \cite{DBLP:conf/ictir/CarteretteBZ15} by enriching test collections with simulated user interactions. Their outlined interaction sequences included the simulation of queries, clicks, dwell times, and session abandonment. Even though they implemented some specific simulators as part of their experiments, they intended to provide a general framework that covers all elements of user interactions which can also be implemented with various methods. More recently, similar frameworks were introduced by Pääkkönen et al. as Common Interaction Model \cite{DBLP:journals/ir/PaakkonenKKAMJ17} and Zhang et al. \cite{DBLP:conf/ictir/ZhangLZ17}.

Most of the current methods for query simulations follow a two-stage approach including the \textit{term candidate generation} and the \textit{query modification strategy}. Usually, the term candidates are derived from a language model. Jordan et al. introduced Controlled Query Generation (CQG) \cite{DBLP:conf/jcdl/JordanWG06} that exploited the relative entropy of a language model for query term generation. Azzopardi et al. applied CQG when generating queries for known-item search \cite{DBLP:conf/sigir/AzzopardiR06,DBLP:conf/sigir/AzzopardiRB07}. In a similar vein, Berendsen et al. used annotations to group documents \cite{DBLP:conf/clef/BerendsenTRM12} and Huurnik et al. simulated queries for purchased items \cite{DBLP:conf/clef/HuurninkHRB10}. When query term candidates are available, there exist some commonly used query modification strategies \cite{DBLP:conf/sigir/Azzopardi11,DBLP:conf/sigir/BaskayaKJ12,DBLP:conf/jcdl/JordanWG06,DBLP:conf/airs/KeskustaloJPSL09}, which were also applied in follow-up studies \cite{DBLP:conf/cikm/MaxwellA16,DBLP:conf/sigir/MaxwellA16,DBLP:conf/ecir/VerberneSJK15} and followed a principled way resulting in controlled query reformulation patterns (cf. Section~\ref{subsec:query_sim}).

If large-scale user logs are available, different approaches propose, for instance, learning to rewrite queries \cite{DBLP:conf/cikm/HeTOKYC16}, model syntactic and semantic changes between query reformulations \cite{DBLP:conf/sigir/HerdagdelenCMHHRA10}, or replace old query terms with new phrases with the help of the point-wise mutual information \cite{DBLP:conf/www/JonesRMG06}. In contrast to these examples, the query simulations analyzed in this study do not rely on large-scale user logs but make use of test collections, i.e., topics and relevance judgments. 

As part of follow-up studies related to the TREC Session Track, Guan et al. improved session search results by introducing the Query Change Model (QCM) \cite{DBLP:conf/sigir/GuanZY13,DBLP:journals/tois/YangGZ15} according to which the session search is modeled as a Markov Decision Process that considers transitions between states, i.e., queries and other interactions, to improve search results for query reformulations. Van Gysel et al. found that QCM is especially effective for longer sessions while being on par with term-frequency based approaches for shorter sessions \cite{DBLP:conf/ictir/GyselKR16}. Our query simulation method draws inspiration from QCM, but generates queries instead of improving retrieval results throughout a session.

Simulated UQVs contribute to more diverse and more realistic user-oriented directives as part of system evaluations. Besides the actual simulation of session search, applications for simulated queries are manifold. For instance, UQVs enhance the pooling process \cite{DBLP:conf/cikm/MoffatSTB15}, make rank fusion approaches possible \cite{DBLP:journals/tois/BenhamMMC19}, are used for query performance prediction \cite{DBLP:conf/ecir/FaggioliZCFS21}, or assist users with query suggestions that improve the recall \cite{DBLP:conf/ecir/VerberneSK14}. In this work, we compare simulated to real UQVs.

\section{Approach}
In this section, we introduce the analyzed approaches for query simulations (\ref{subsec:query_sim}) featuring conventional methods of term candidate generation and query modification strategies and our new method. Furthermore, the evaluation framework (\ref{subsec:eval}) and details about the datasets and implementations (\ref{subsec:data}) are described. 

\subsection{Query simulation}
\label{subsec:query_sim}

\subsubsection{Term candidate generation}
Simulating queries based on topics of test collections most likely complies with exploitation search tasks \cite{DBLP:conf/chiir/LiuSS20}, where users normally have a very concrete understanding of their information needs. Provided that real users have read the topic, they are very likely to include key terms of the topic texts when formulating queries. As a simplified implementation, the \textit{TREC Topic Searcher} (TTS) considers only terms of the set $T_{\mathrm{topic}}=\{t_1,..., t_n\}$ composed of the topic's title, description and narrative with $t_1,..., t_n$ being the term sequence in the concatenated text. For upper bound performance estimates, we simulate a \textit{Known-item Searcher} (KIS). Here, we assume the simulated users to be familiar with the document collection. When reading the topics, they recall key terms of the relevant documents in the collection and use these as their query terms. In this case, the term candidates $T_{\mathrm{rel}}=\{t_1,...,t_n\}$ are derived from a language model based on CQG by Jordan et al. \cite{DBLP:conf/jcdl/JordanWG06} according to $ P(t | D_{\mathrm{rel}}) = (1 - \lambda)P_{\mathrm{topic}}(t | D_{\mathrm{rel}}) + \lambda P_{\mathrm{background}}(t)$, where the topic model $P_{\mathrm{topic}}(t | D_{\mathrm{rel}})$ is made from the relevant documents $D_{\mathrm{rel}}$ for a given topic, while the background model $P_{\mathrm{background}}(t)$ is derived from the vocabulary of the entire corpus. $\lambda$ is used to model the influence of the background model, and it is set to $0.4$ to be consistent with previous work \cite{DBLP:conf/sigir/Cronen-TownsendZC02,DBLP:conf/jcdl/JordanWG06}. In this case, $t_1,..., t_n$ are ordered by the decreasing term probabilities of the underlying language model.

\subsubsection{Query modification strategy}
We make use of the query generation techniques proposed by Baskaya et al. \cite{DBLP:conf/sigir/BaskayaKJ12}, that were also used in previous simulation studies \cite{DBLP:conf/jcdl/JordanWG06,DBLP:conf/cikm/MaxwellA16,DBLP:conf/sigir/MaxwellA16,DBLP:journals/ir/PaakkonenKKAMJ17,DBLP:conf/ecir/VerberneSJK15}. More specifically, the following strategies are considered and used in combination with the term candidates of $T_{\mathrm{topic}}$ and $T_{\mathrm{rel}}$: the strategy S1 outputs single term queries $q_i$ following the ordering of term candidates ($q_1 = \{t_1\}; q_2 = \{t_2\}; q_3 = \{t_3\}; ...$); S2 keeps the first candidate term fixed and composes query strings by replacing the second term for reformulations ($q_1 = \{t_1, t_2\}; q_2 = \{t_1, t_3\}; q_3 = \{t_1, t_4\};...$); S$2^\prime$ is similar to S2, but keeps two candidate terms fixed ($q_1 = \{t_1, t_2, t_3\}; q_2 = \{t_1, t_2, t_4\}; q_3 = \{t_1, t_2, t_5\};...$); S3 starts with a single term query and incrementally adds query terms for reformulations ($q_1 = \{t_1\}; q_2 = \{t_1, t_2\}; q_3 = \{t_1, t_2, t_3\};...$); S$3^\prime$ is similar to S3, but starts with two candidate terms ($q_1 = \{t_1, t_2, t_3\}; q_2 = \{t_1, t_2, t_3, t_4\}; q_3 = \{t_1, t_2, t_3, t_4, t_5\};...$). In total, we analyze ten different query simulators that result from the two term candidate generators that are combined with five query modification strategies, denoted as TTS\textsubscript{S1-S$3^\prime$} and KIS\textsubscript{S1-S$3^\prime$}, respectively. We hypothesize, that the system performance of real queries should range somewhere between those queries of the naive approach of TTS and those queries of KIS. 

\subsubsection{Controlled Query Generation combined with Query Change Model}
Compared to the previous query simulators, this approach adds an additional scoring stage for the generated query string candidates. These candidates are generated by considering every possible combination of n-grams made from a term set. The corresponding terms are either taken from $T_{\mathrm{rel}}$ or $T_{\mathrm{topic+rel}}= (T_{\mathrm{topic}} \cap T_{\mathrm{rel}}) \cup (T_{\mathrm{rel}} \setminus T_{\mathrm{topic}})_k$, whereas $(T_{\mathrm{topic}} \cap T_{\mathrm{rel}})$ contains topic terms in $T_{\mathrm{rel}}$ and $(T_{\mathrm{rel}} \setminus T_{\mathrm{topic}})_k$ denotes the top $k$ terms of $T_{\mathrm{rel}}$ that are not in the topic text. In this regard, $k$ models the user's vocabulary and domain knowledge. Having a set of different query string candidates, we rank the queries by $\frac{\sum_{j=1}^{|q|} \Theta_j}{|q|}$, which is the sum over all query terms normalized by the query length $|q|$, whereas $\Theta_j$ is a term-dependent score inspired by QCM \cite{DBLP:conf/sigir/GuanZY13,DBLP:journals/tois/YangGZ15} and is implemented as follows.

\begin{equation}
  \Theta_j = 
  \begin{cases}
    \alpha (1 - P (t_j | D_{\mathrm{rel}})), & t_j \in q_{\mathrm{title}} \\
    1 - \beta P (t_j | D_{\mathrm{rel}}), & t_j \in + \Delta q \land t_j \in T_{\mathrm{topic}} \\
    \epsilon \ \mathrm{idf}(t_j), & t_j \in + \Delta q \land t_j \notin T_{\mathrm{topic}} \\
    - \delta  P (t_j | D_{\mathrm{rel}}), & t_j \in - \Delta q \\
  \end{cases}
\end{equation}

whereas $q_{\mathrm{title}}$ is the set of topic title terms and $+/- \Delta q$ denotes added or removed terms of a query reformulation that is made in reference to the previously simulated query, except for the first query formulation $q_1$ for which the topic title is used as a reference. In our experiments, we analyze 3-,4-,5-gram term candidates and analyze three different parametrizations of the simulators, which are defined as follows. First, we analyze the strategy S4 ($\alpha=2.2,\beta=0.2,\epsilon=0.05,\delta=0.6$), which tends to prefer topic terms and mostly keeps terms of previous queries. Second, we analyze the strategy S$4^\prime$ ($\alpha=2.2,\beta=0.2,\epsilon=0.25,\delta=0.1$), which mostly keeps terms of previous queries, but tends to include terms that are not in the topic text. Finally, we analyze the strategy S$4^{\prime\prime}$ ($\alpha=0.2,\beta=0.2,\epsilon=0.025,\delta=0.5$), which tends to stick to the topic terms, but does not necessarily keep terms of previous query formulations. In sum, we analyze six different instantiations of these simulators, which are either based on $T_{\mathrm{rel}}$ (denoted as KIS\textsubscript{S4-S$4^{\prime\prime}$}), or based on $T_{\mathrm{topic+rel}}$ with $k=4$ (denoted as TTS\textsubscript{S4-S$4^{\prime\prime}$}).

\subsection{Evaluation framework}
\label{subsec:eval}

In the following, we outline our evaluation framework used to validate the simulations in reference to real queries in different aspects. It includes evaluation of the average retrieval performance, shared task utility, effort and effect, and query term similarity between simulated and real queries.

\subsubsection{Retrieval performance} 
As shown by Tague and Nelson, simulated queries fall behind real queries in terms of retrieval performance \cite{DBLP:conf/sigir/TagueN81}. For this reason, we evaluate the \textit{Average Retrieval Performance} (ARP) as it is common practice in system-oriented IR experiments. The ARP is determined by the average of a measure over all topics in a test collection. Beyond comparing the averaged means of different queries, we propose a more in-depth analysis of the topic score distributions. Recently, the \textit{Root Mean Square Error} (RMSE) and \textit{paired t-tests} were introduced as reproducibility measures \cite{DBLP:conf/sigir/Breuer0FMSSS20}. The RMSE measures the closeness between the topic score distributions, and low errors indicate a good reproduction. When using t-tests as a reproducibility measure, low p-values result from diverging score distributions and indicate a higher probability of failing the reproduction. 

\subsubsection{Shared task utility} 
According to Huurnik et al. \cite{DBLP:conf/clef/HuurninkHRB10}, the ARP of the simulated queries alone is not an appropriate indicator of how well the simulations resemble the real queries since useful query simulators should identify the best system. As proposed by Huurnik et al., we analyze how the simulated queries reproduce \textit{relative system orderings} by comparing them with the help of Kendall's $\tau$ as it is common practice as part of shared task evaluations \cite{DBLP:conf/sigir/Voorhees98}. We compare the simulated and real queries by determining how well the ordering of systems with different parametrizations (and different retrieval performance) can be reproduced by simulated queries. 

\subsubsection{Effort and effect} 
In order to account for a more user-oriented evaluation, we simulate sessions and evaluate them with regards to the effort (number of queries) that has to be made and the resulting effects (cumulated gain). First, we simulate sessions using ten simulated queries and an increasing number of documents per query and evaluate the results by the \textit{sDCG} measure \cite{DBLP:conf/ecir/JarvelinPDN08}, whereas the cumulated gain is discounted for each result and query. Second, we evaluate the simulation quality from another more economical point of view. Azzopardi applies economic theory to the retrieval process \cite{DBLP:conf/sigir/Azzopardi11} and demonstrates that for a pre-defined level of cumulated gain, query reformulations can be compensated by browsing depth (or vice versa browsing depth by more query reformulations). Furthermore, he illustrates this relationship with isoquants - a visualization technique used in microeconomics. Thus, we evaluate the closeness between isoquants of simulated and real queries by the \textit{Mean Squared Logarithmic Error} (MSLE).

\subsubsection{Query term similarity} 
It is not the primary goal of this study to simulate query strings with exact term matches.  Instead, simulated UQVs should result in diverse query strings for a fixed information need (topic). Nonetheless, it is worth analyzing the term overlap between the simulated and real queries. As Liu et al. \cite{DBLP:conf/ictir/LiuC0KC19} or Mackenzie and Moffat \cite{DBLP:conf/ictir/MackenzieM21} propose, we determine the Jaccard similarity between the sets of unique terms made from the query reformulations. When compared with the other evaluations, the term similarities add more insights about the simulated UQVs. For instance, if it is possible to simulate query reformulations that adequately relate to the properties of real queries, but with other terms.

\subsection{Datasets and implementation details}
\label{subsec:data}
In our experimental setup, we use the user query variant (UQV) dataset provided by Benham and Culpepper~\cite{DBLP:conf/adcs/BenhamC17}\footnote{\url{https://culpepper.io/publications/robust-uqv.txt.gz}}. Given the topic texts, eight users formulated up to ten query variants for each topic. Each user formulated at least one query for each topic, and the fifth user (denoted as UQV\textsubscript{5}) formulated ten queries for each topic. More details about the query collection process are provided by Benham et al. \cite{DBLP:conf/trec/BenhamGMDCSMC17}. Accordingly, we evaluate the system runs with The New York Times Annotated Corpus and the topics of TREC Common Core 2017 \cite{DBLP:conf/trec/AllanHKLGV17}.
As part of our experiments, we exploit the interactive search possibilities of the Pyserini toolkit \cite{DBLP:conf/sigir/LinMLYPN21}. We index the Core17 test collection with the help of Anserini \cite{DBLP:journals/jdiq/YangFL18} and the default indexing options as provided in the regression guide\footnote{\url{https://github.com/castorini/anserini/blob/master/docs/regressions-core17.md}}. Unless stated otherwise, all results were retrieved with the BM25 method and Anserini's default parameters ($b=0.4$, $k=0.9$). We evaluate the results with the \texttt{repro\_eval} toolkit \cite{DBLP:conf/ecir/BreuerFMS21} that is a dedicated reproducibility framework featuring bindings to \texttt{trec\_eval} measures. The source code of the experiments and the simulated queries are available in a public GitHub repository\footnote{\faicon{github} \url{https://github.com/irgroup/ecir2022-uqv-sim}}.

\section{Experimental results}
\label{sec:experiments}

\subsubsection{Retrieval performance}
Regarding \textbf{RQ1}, we validate the retrieval performance of real (UQV) and simulated (TTS/KIS) queries. Table \ref{tab:arp} shows the ARP including nDCG and AP scores that are determined by averaging results with 1000 documents per topic and P@10 scores over \textit{all} queries, the \textit{first}\footnote{S1 and S3, as well as S2 and S$3^\prime$, do not differ when averaging over the first queries.}, or the \textit{best} query of a topic. Our assumptions are confirmed. The retrieval performance of real queries ranges between that of the TTS\textsubscript{S1-S$3^\prime$} and KIS\textsubscript{S1-S$3^\prime$} simulators. Especially, the performance of the TTS\textsubscript{S1-S$3^\prime$} queries stays below that of real queries. For instance, the average nDCG scores of the UQV queries range between $0.3787$ and $0.4980$, whereas the maximum score of the TTS\textsubscript{S1-S$3^\prime$} queries is $0.3499$ and the nDCG scores of KIS\textsubscript{S$2^\prime$-S$3^\prime$} lie above those of UQV. Similarly, the nDCG scores averaged over the first UQV queries reach $0.3979$ at a minimum, whereas the maximum score of the TTS\textsubscript{S1-S$3^\prime$} queries is $0.3895$. When averaging over the best queries, most nDCG scores of TTS fall into the range of real queries, but there is also a higher probability of finding a good performing query since more TTS than UQV queries are available. Except for single term queries (S1), all KIS scores outperform the UQV queries when averaging over the best queries. With regard to the simulated queries based on the TTS\textsubscript{S4-S$4^{\prime\prime}$} approach, most of the nDCG, P@10, and AP scores fall into the range of the real queries, while KIS\textsubscript{S4-S$4^{\prime\prime}$} queries outperform UQV queries. Thus, we have a specific focus on TTS\textsubscript{S4-S$4^{\prime\prime}$}. 

\begin{figure}[!ht]
  \centering
  \includegraphics[width=0.325\textwidth]{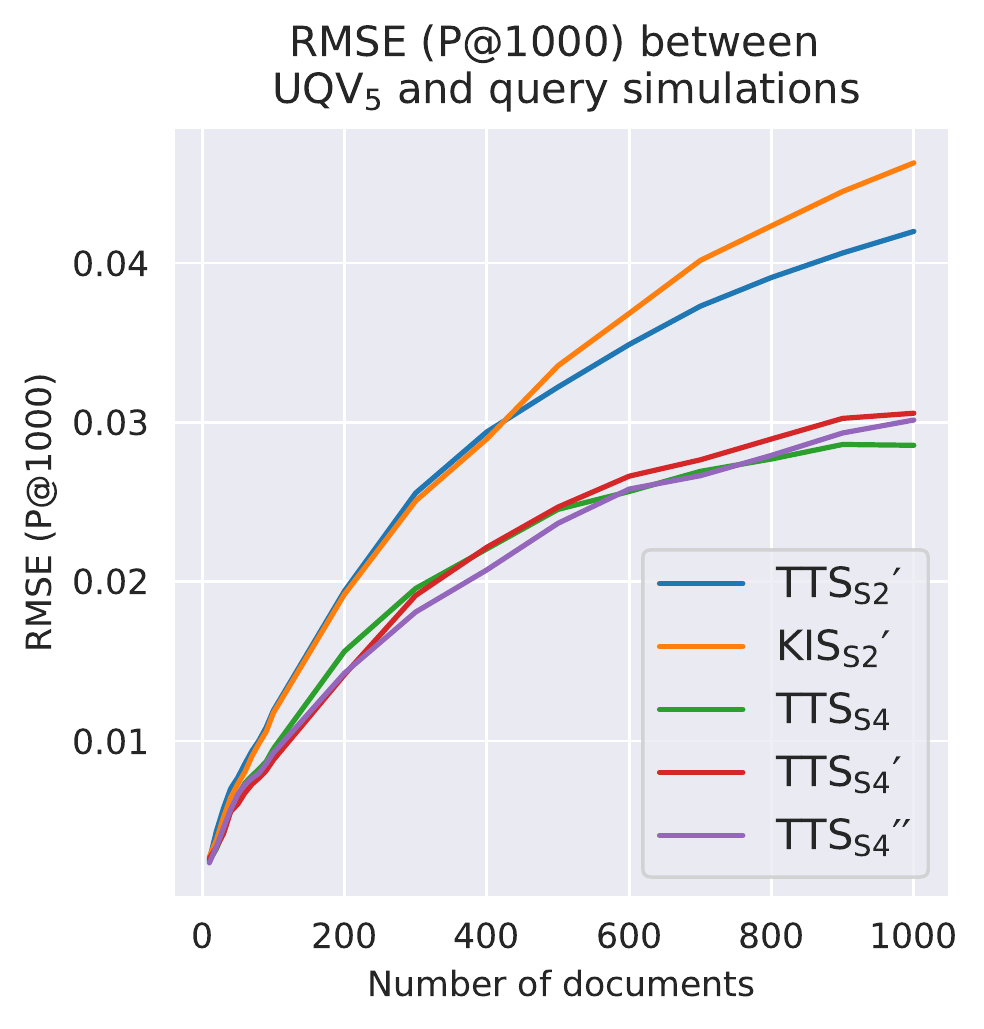}
  \includegraphics[width=0.325\textwidth]{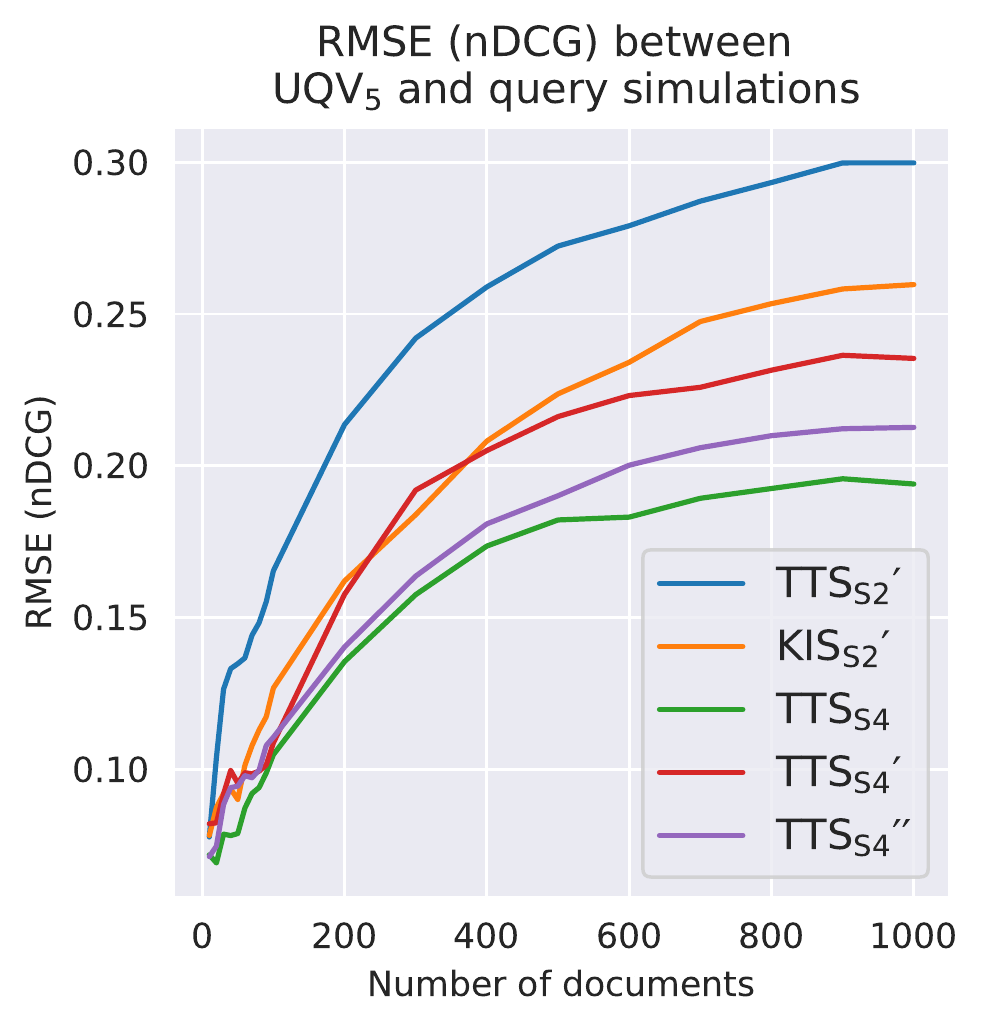}
  \includegraphics[width=0.325\textwidth]{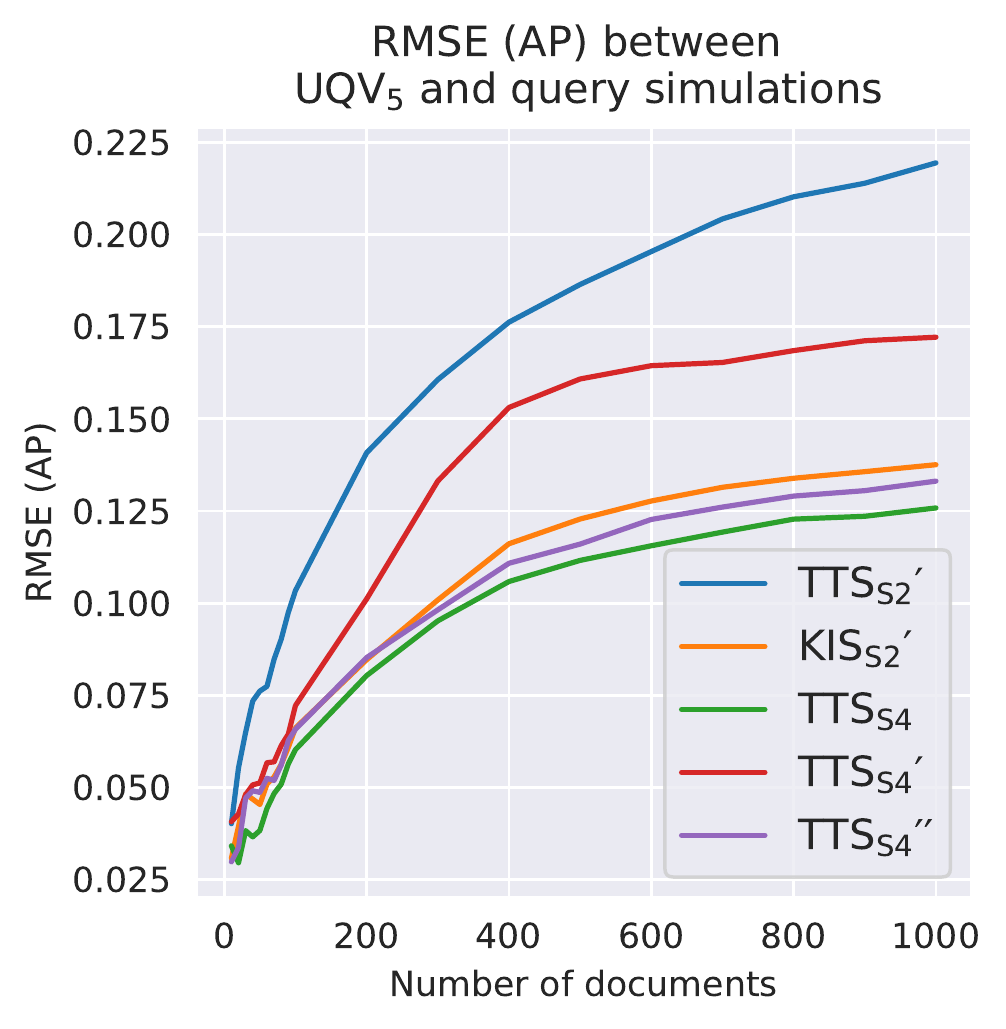}
  \caption{RMSE instantiated with P@1000, nDCG, and AP}
  \label{fig:nrmse_p1000_ndcg_map}
\end{figure}

Figure \ref{fig:rmse_s1s5} shows the RMSE\textsubscript{nDCG} between queries with conventional query modification strategies (TTS\textsubscript{S1-S$3^\prime$}/KIS\textsubscript{S1-S$3^\prime$}) and the real queries (UQV). Especially for the TTS queries, the strategy S$2^\prime$ has the lowest RMSE scores and acceptable scores for the KIS queries. In the following experiments, we primarily use the strategy S$2^\prime$ for both the TTS and KIS queries since their term length complies with the typical length of real queries \cite{DBLP:journals/jasis/JansenBS09} and they serve as estimates of lower and upper bound retrieval performance. Additionally, we evaluate the TTS\textsubscript{S4-S$4^{\prime\prime}$} queries with the help of the RMSE and simulations in reference to the ten queries per topic of UQV\textsubscript{5}. For each query reformulation, 100 documents are retrieved and contribute to the final ranking list of a topic if a previous query has not retrieved them. Figure \ref{fig:nrmse_p1000_ndcg_map} shows the RMSE instantiated with P@1000, nDCG, and AP along with an increasing number of documents retrieved with ten queries. For all measures, the error increases when more documents per query are retrieved. With regard to P@1000 and nDCG, the TTS\textsubscript{S$2^\prime$} and KIS\textsubscript{S$2^\prime$} queries have the largest error, while KIS\textsubscript{S$2^\prime$} has a lower RMSE\textsubscript{AP} than TTS\textsubscript{S$4^\prime$}. For all measures, 
the TTS\textsubscript{S4-S$4^{\prime\prime}$} queries have the lowest error, which means they are the best approximation of UQV\textsubscript{5} among all analyzed query simulations.

Finally, we compare the topic score distributions of the simulated queries and all UQV queries by paired t-tests\footnote{Applying the Bonferroni correction adjusts the alpha level to $\alpha=\frac{0.05}{64}\approx0.0008$ (considering eight users and eight query simulators for an alpha level of 0.05).}. Since some users formulated no more than one query per topic, we limit our evaluations to the first query of each simulator. It means that each of the p-values shown in Figure \ref{fig:ttest} is determined by t-tests with nDCG score distributions that result from 50 UQV and 50 simulated queries. The TTS\textsubscript{S$2^\prime$} queries have the highest p-values when compared with UQV\textsubscript{\{2,3,8\}}. These results align with the ARP scores reported in Table \ref{tab:arp}. The nDCG scores of UQV\textsubscript{2} (0.4096), UQV\textsubscript{3} (0.3979), and UQV\textsubscript{8} (0.4046) are the most similar to the nDCG score of TTS\textsubscript{S3} (0.3895) in comparison to other simulators. In contrast, the p-values of KIS\textsubscript{S$2^\prime$} queries are low for all UQV queries, which complies with the ARP scores in Table \ref{tab:arp}. The KIS\textsubscript{S$2^\prime$} scores averaged over the first queries are substantially higher compared to the UQV scores (e.g., nDCG(KIS\textsubscript{S$2^\prime$})=0.5474  compared to the best UQV query with nDCG(UQV\textsubscript{7})=0.4980). The UQV\textsubscript{\{1,4,5,6,8\}} queries have comparably higher p-values with the TTS\textsubscript{\{S4,S$4^{\prime\prime}$\}} queries which align with similar ARP scores. Interestingly, the t-test with UQV\textsubscript{7} and TTS\textsubscript{S$4^{\prime}$} results in the highest overall p-value of 0.9901 and similarly high p-values with KIS\textsubscript{S4-S$4^{\prime\prime}$}. This lets us assume that the corresponding user of the UQV\textsubscript{7} queries diverged from the terms in the topic texts and had some prior knowledge about adequate queries for at least some of the topics. In sum, not only the ARP can be reproduced with the simulated TTS\textsubscript{S4-S$4^{\prime\prime}$} and KIS\textsubscript{S4-S$4^{\prime\prime}$} queries, but also statistical properties of the topic score distributions.

\subsubsection{Shared task utility}
\begin{figure}[t]
  \centering
  \includegraphics[width=0.45\textwidth]{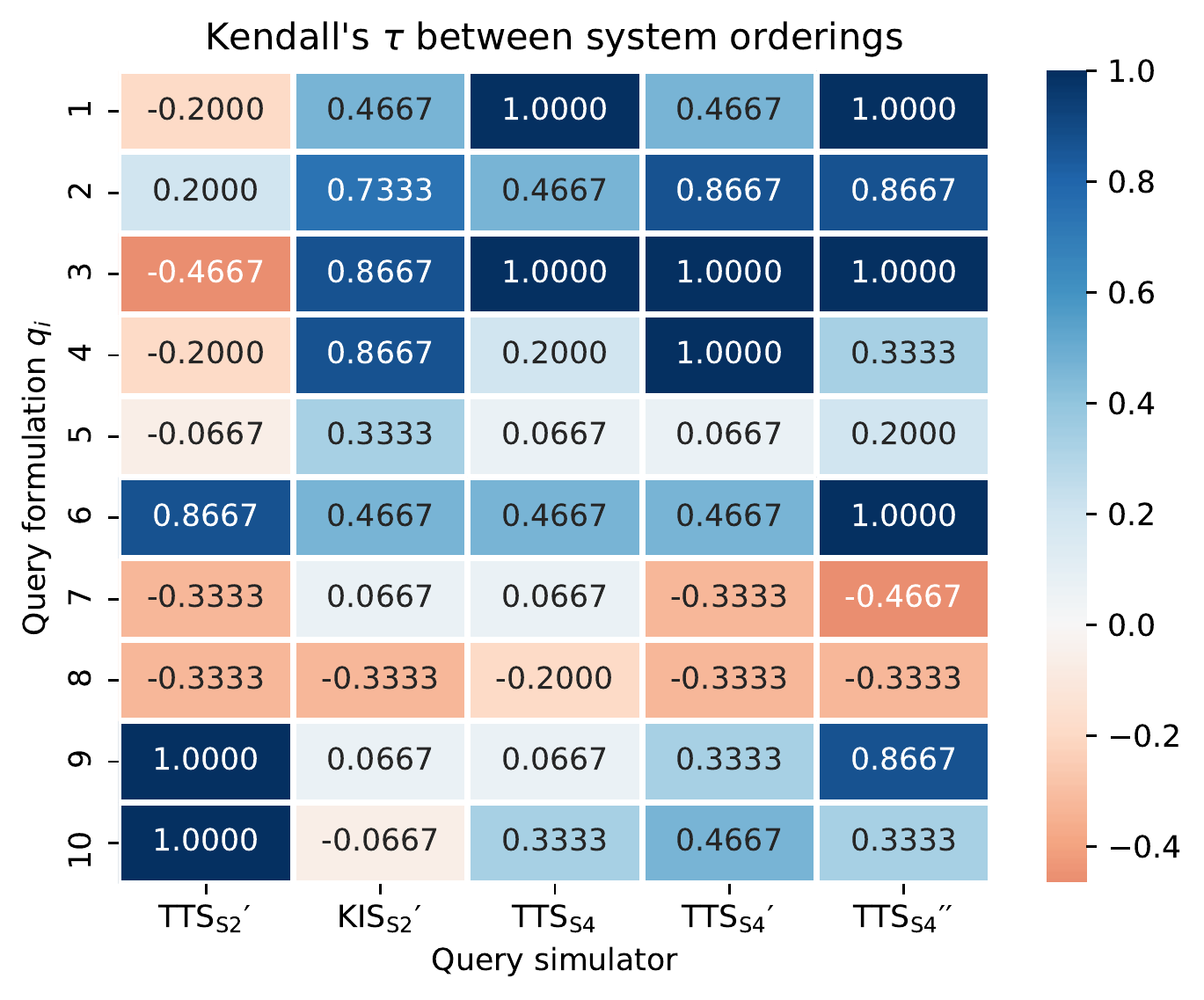}
  \includegraphics[width=0.45\textwidth]{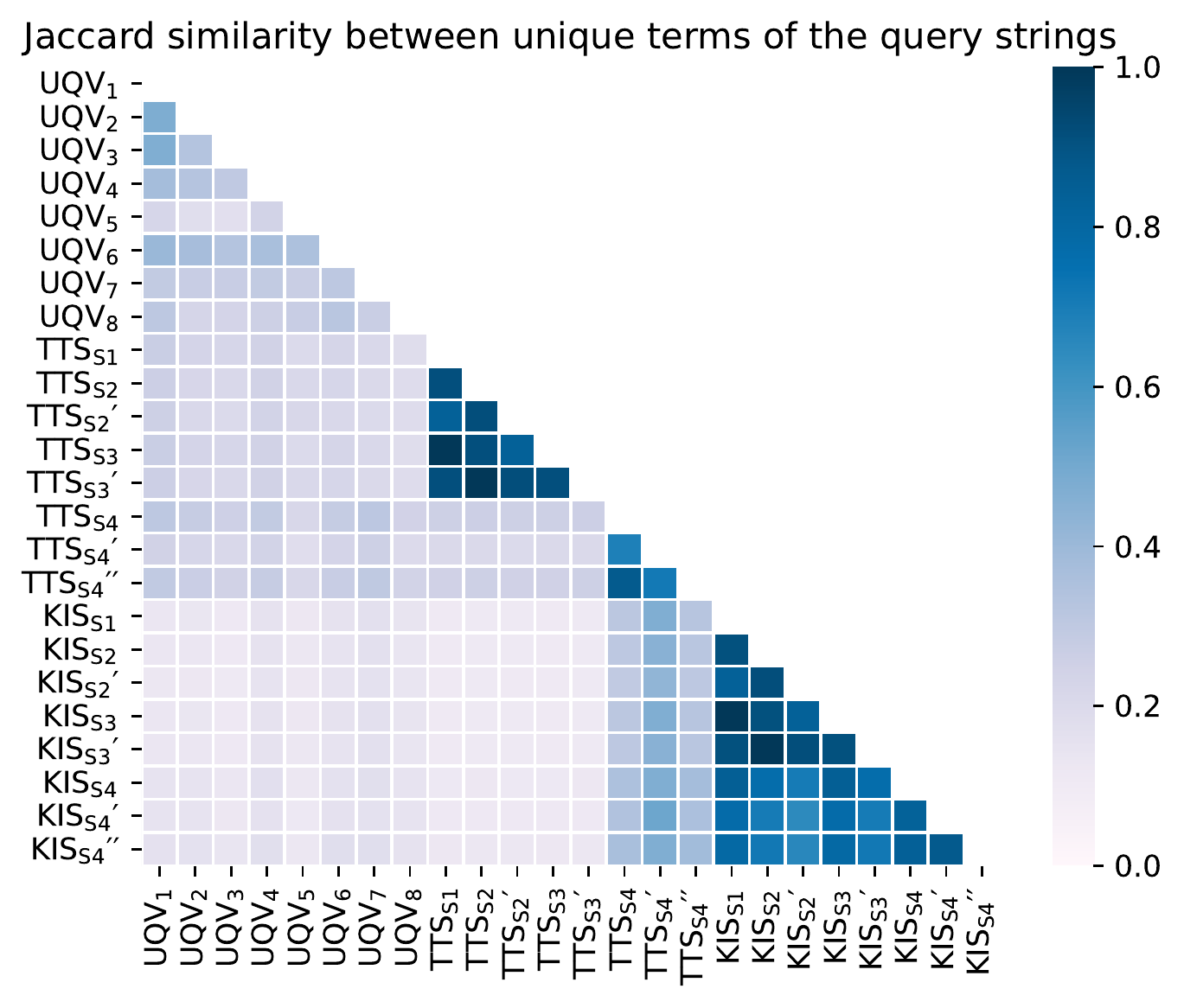}
  \caption{Kendall's $\tau$ between system orderings of query reformulations in reference to UQV\textsubscript{5} (left). Jaccard similarity between unique terms of the queries (right).}
  \label{fig:kt_jacc}
\end{figure}

Regarding \textbf{RQ2}, we validate to which degree the simulated queries reproduce properties of the real queries in several regards. First, we evaluate if the simulated queries can preserve the relative system orderings. To be consistent with Huurnik et al., we evaluate five systems and different parametrizations ($\mu=50, 250, 500, 1250, 2500, 5000$) of the query likelihood model with Dirichlet smoothing (QLD) \cite{DBLP:conf/sigir/ZhaiL01}, but other retrieval methods and variations thereof can be reasonable as well. For each query formulation $q_i$, we determine the correlation by Kendall's $\tau$ averaged over all topics (cf. Figure \ref{fig:kt_jacc} (left)) in comparison to the UQV\textsubscript{5} queries. The TTS\textsubscript{S$2^\prime$} queries do not preserve the relative system ordering. Especially for the first five query reformulations, there is a low correlation with the relative system orderings of the real queries. Interestingly, the KIS\textsubscript{S$2^\prime$} queries result in acceptable Kendall's $\tau$ scores \cite{DBLP:conf/sigir/Voorhees98}, while the scores beyond the sixth query formulation show low correlations. Similarly, the TTS\textsubscript{S4-S$4^{\prime\prime}$} queries correlate with the system orderings of UQV\textsubscript{5} queries fairly well, even reaching the maximum score of 1.0. Beyond the sixth query reformulation, the correlation falls off. While it is out of this study's scope to reach any deﬁnitive conclusions, we assume that this is related to query drifts - an issue that is also known from term expansions as part of pseudo-relevance feedback \cite{DBLP:journals/jd/CroftH79,DBLP:journals/ker/RuthvenL03}.

\subsubsection{Effort and effect}
\begin{figure}[!ht]
  \begin{minipage}{\linewidth}
    \includegraphics[width=0.325\textwidth]{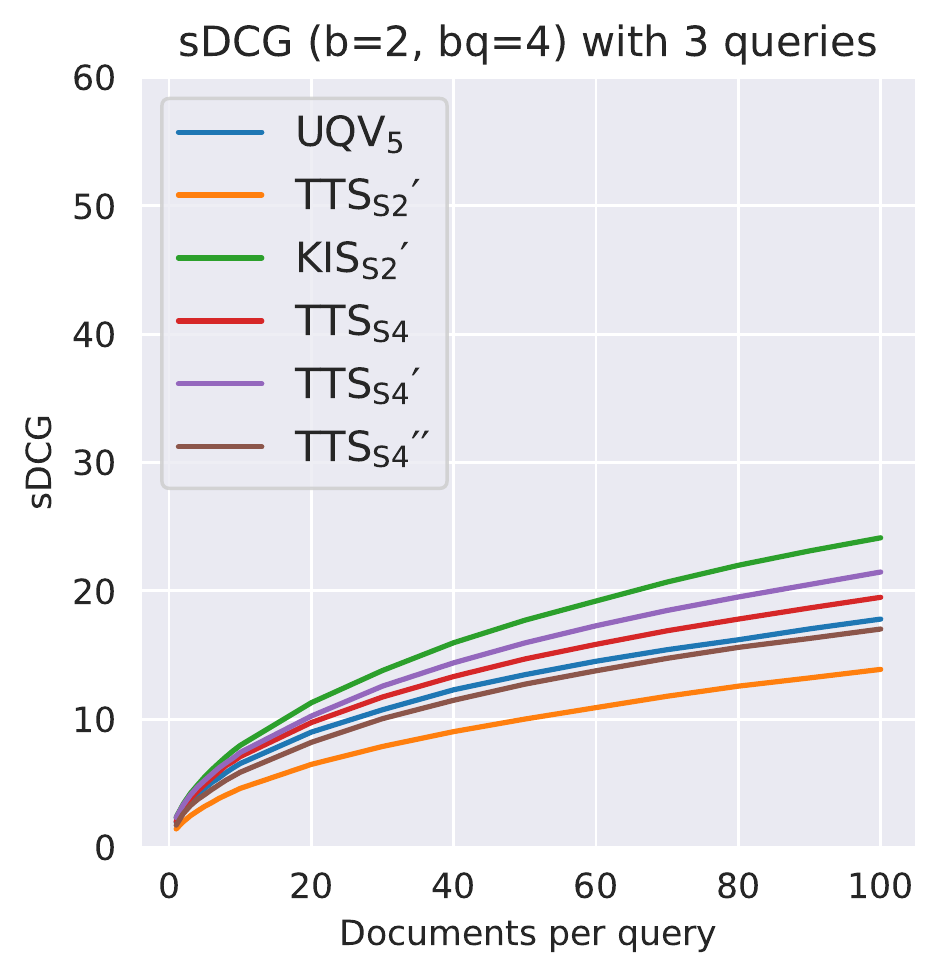}
    \includegraphics[width=0.325\textwidth]{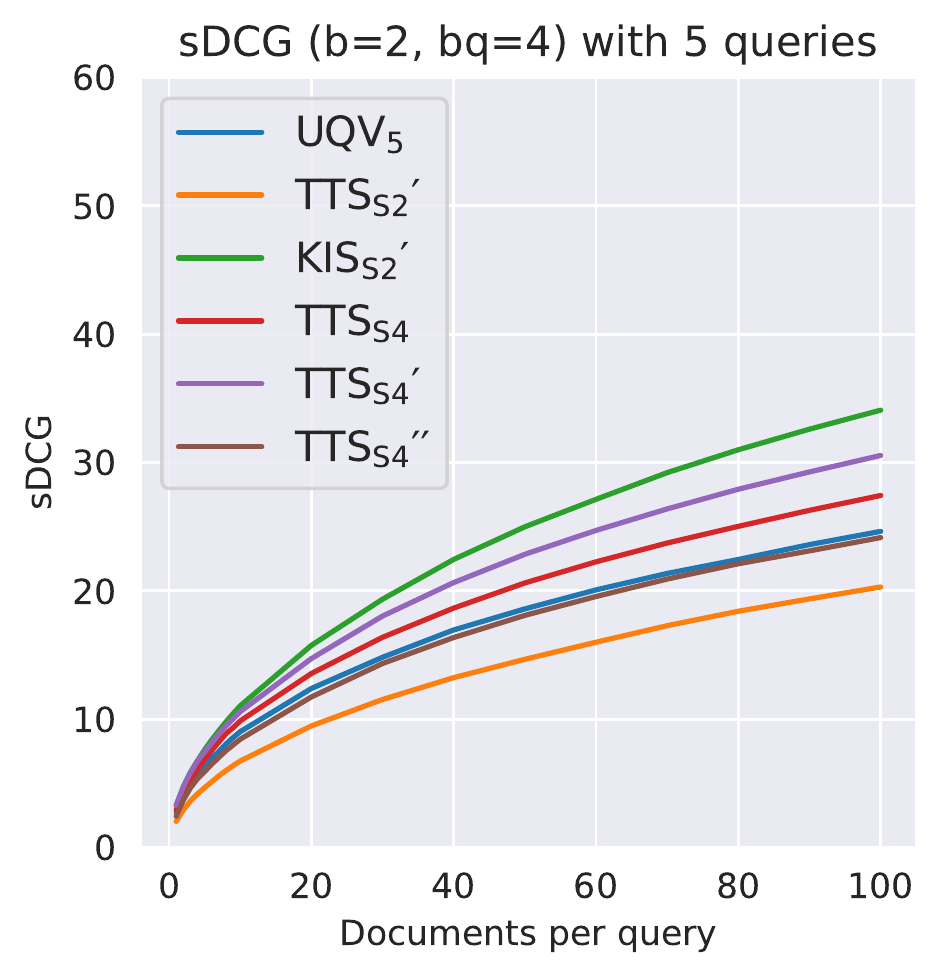}
    \includegraphics[width=0.325\textwidth]{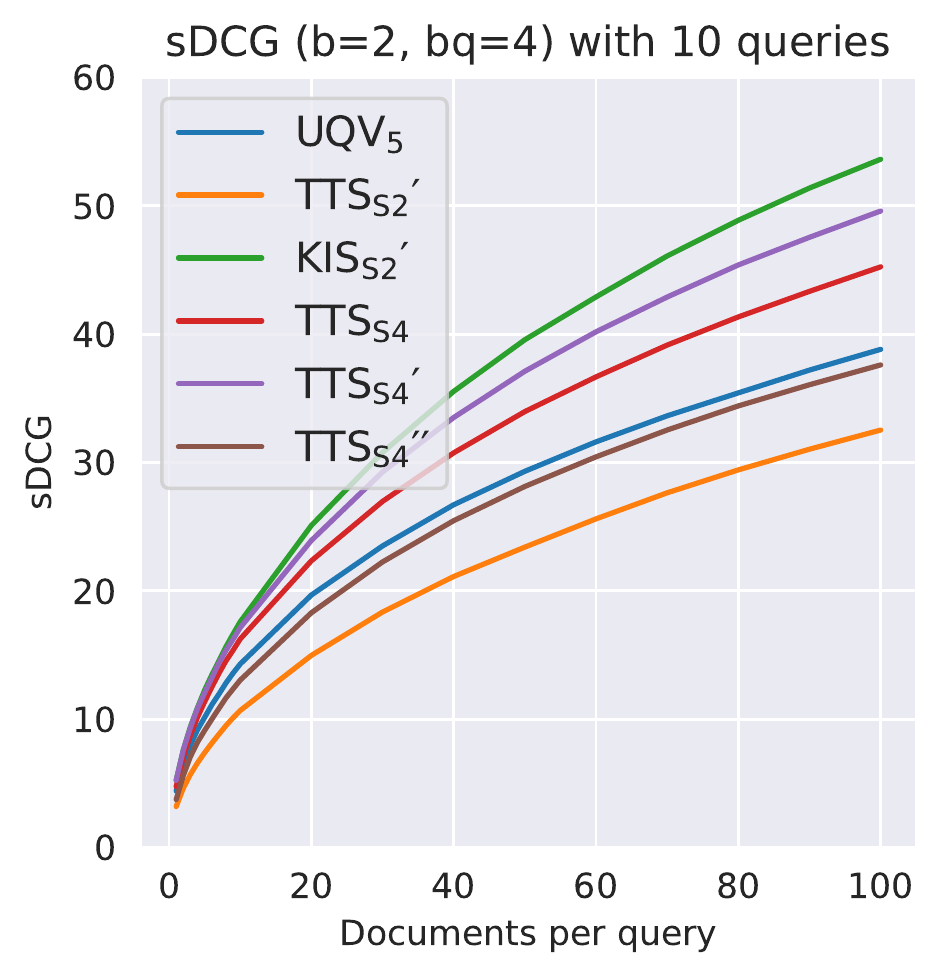}
  \end{minipage}
  \begin{minipage}{0.69\linewidth}
    \includegraphics[width=0.475\textwidth]{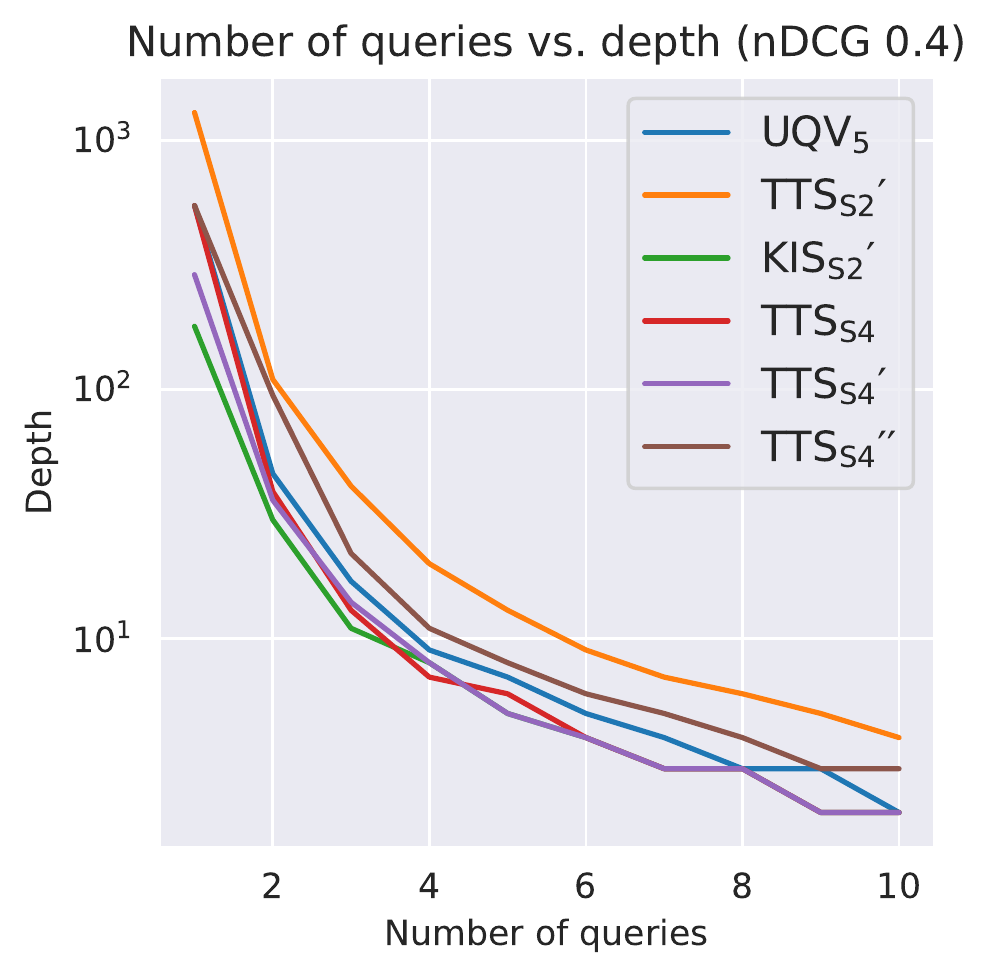}
    \includegraphics[width=0.475\textwidth]{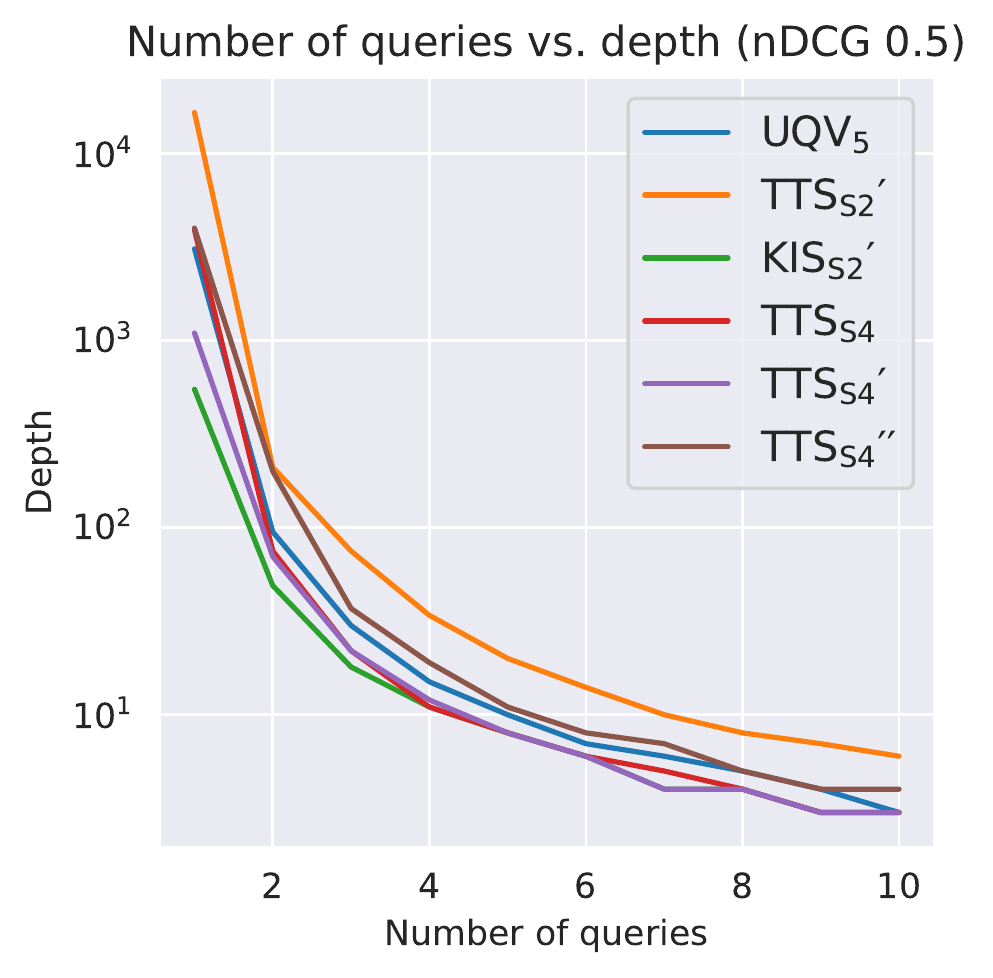}
  \end{minipage}
  \begin{minipage}{0.3\linewidth}
    \resizebox{\columnwidth}{!}{
      \begin{tabular}{h|z|y|z}
        \multicolumn{4}{c}{MSLE} \\
        \vspace{1mm}\\
        \toprule
        \multicolumn{1}{l}{nDCG} & \multicolumn{1}{c}{0.3} & \multicolumn{1}{c}{0.4} & \multicolumn{1}{c}{0.5} \\
        \midrule
        \multicolumn{1}{l}{TTS\textsubscript{S$2^\prime$}} & 0.3371 & 0.4279 & 0.6568 \\
        \multicolumn{1}{l}{KIS\textsubscript{S$2^\prime$}} & 0.0949 & 0.1837 & 0.3987 \\
        \multicolumn{1}{l}{TTS\textsubscript{S4}} & 0.0059 & 0.0323 & 0.0444 \\
        \multicolumn{1}{l}{TTS\textsubscript{S$4^\prime$}} & 0.0509 & 0.0758 & 0.1550 \\
        \multicolumn{1}{l}{TTS\textsubscript{S$4^{\prime\prime}$}} & 0.0713 & 0.0807 & 0.0791\\
        \bottomrule
      \end{tabular}
    }
    \vspace{10pt}
  \end{minipage}
  \captionof{figure}{Simulations with 3, 5, or 10 queries per session evaluated by sDCG (top). Isoquants and MSLE between simulations and UQV\textsubscript{5} with fixed nDCG (bottom).}
  \label{fig:economic}
\end{figure}

Since most of the experiments validated single queries only, we simulate search sessions and evaluate these by sDCG (instantiated with b=2, bq=4). We compare sessions with 3, 5, or 10 queries and an increasing number of documents per query. Figure \ref{fig:economic} (top) compares the queries of UQV\textsubscript{5} (made by a single user \cite{DBLP:conf/trec/BenhamGMDCSMC17}) to ten simulated queries of TTS\textsubscript{S$2^\prime$}, KIS\textsubscript{S$2^\prime$}, and TTS\textsubscript{S4-S$4^{\prime\prime}$}. As expected, the cumulative gain increases faster when more queries per session are used. Likewise, the TTS\textsubscript{S$2^\prime$} and KIS\textsubscript{S$2^\prime$} queries deliver lower and upper bound limits, respectively. In between, there are the cumulative gains by the UQV\textsubscript{5} and TTS\textsubscript{S4-S$4^{\prime\prime}$} queries. These results show that it is possible to fine-tune and to reproduce the cumulative gain close to that of real queries, in this particular case with TTS\textsubscript{S$4^{\prime\prime}$}.

Figure \ref{fig:economic} (bottom) shows the isoquants and illustrates how many documents have to be examined by a simulated user to reach pre-defined levels of nDCG (0.3, 0.4, 0.5). More queries compensate browsing depth, and as expected, the least documents have to be examined with KIS\textsubscript{S$2^\prime$} queries and the most with TTS\textsubscript{S$2^\prime$} queries. The TTS\textsubscript{S$2^\prime$} isoquants lie above the others, which can be explained by the poorer retrieval performance as already shown in Table \ref{tab:arp}. As shown by the MSLE, the TTS\textsubscript{S4} isoquant has the lowest error for all values of nDCG. Again, we see a better approximation of the UQV\textsubscript{5} isoquant with the TTS\textsubscript{S4-S$4^{\prime\prime}$} strategies and that it is possible to reproduce economic properties through parameterizing the query reformulation behavior.

\subsubsection{Query term similarities}
Figure \ref{fig:kt_jacc} (right) shows the Jaccard similarities between the concatenated query strings. More specifically, only normalized unique terms are compared, and depending on the number of available queries for a specific topic, we include an equal number of simulated queries to avoid low Jaccard similarities when less than ten UQV queries are available. As the results show, the highest similarities are between the simulated queries. While the similarities between conventional strategies S1 to S$3^\prime$ and the strategies S4 to S$4^{\prime\prime}$ are rather low for the TTS queries, there are higher similarities for the KIS queries. Compared to UQV and TTS queries, the KIS queries have the lowest similarities, which indicates that descriptive terms of relevant documents are very different from those used in real queries and the topic texts. Interestingly, the UQV\textsubscript{\{2,3,8\}} queries do not have a remarkably high Jaccard similarity with TTS\textsubscript{S$2^\prime$} queries, despite the high p-values that are shown in Figure \ref{fig:ttest}. This shows that it is possible to simulate UQVs with different query terms than in the real queries, but with comparable statistical properties as indicated by the p-values even with the rather naive approach of TTS\textsubscript{S$2^\prime$}. There are slightly higher similarities between KIS queries and the TTS\textsubscript{S4-S$4^{\prime\prime}$} queries. In particular, there is a higher similarity between TTS\textsubscript{S$4^\prime$} and the KIS queries since the simulator is parameterized to diverge from the topic terms. Overall, we conclude that the analyzed simulation methods do not result in query strings that exactly match the terms of real queries in this specific UQV dataset \cite{DBLP:conf/adcs/BenhamC17}.

\section{Discussion}
Referring to our research questions posed earlier, we answer them as follows. 

\subsubsection{RQ1 How do real user queries relate to simulated queries made from topic texts and known-items in terms of retrieval effectiveness?}
It is possible to use the TTS\textsubscript{S1-S$3^\prime$} and KIS\textsubscript{S1-S$3^\prime$} queries, which follow conventional simulation methods, as lower and upper bound estimates between which the retrieval performance of real user query variants (UVQ\textsubscript{1-8}) ranges. Simulations based on our new method (TTS\textsubscript{S$4^{\prime\prime}$}) provide better approximations of real query effectiveness, and the parametrization allows the simulation of different query formulation behaviors and a retrieval performance better resembling real queries.

\subsubsection{RQ2 To which degree do simulated queries reproduce real queries provided that only resources of the test collection are considered for the query simulation?}
Our experiments show that the simulated TTS\textsubscript{S4-S$4^{\prime\prime}$} queries reproduce the real UQV queries reasonably well in several regards. Beyond a similar ARP, they also reproduce statistical properties of the topic score distributions as shown by the RMSE and p-values. Furthermore, it is shown that the simulated queries also reproduce economic aspects of the real queries as evaluated with the sDCG experiments and the isoquants that compare tradeoffs between the number of query reformulations and the browsing depth for a fixed level of gain. Furthermore, when evaluating the shared task utility, the queries of our new parameterized simulation approach preserve the relative system orderings up to the fifth reformulation, while the correlations fall off for later reformulations. We assume that this is related to topic drifts, and further analysis in this direction is required. Finally, even though it is not the primary goal to simulate exact term matches with UQVs, the analysis of the query term similarity showed that there is only a slight overlap between terms of simulated and real queries, and a more dedicated approach is required to reproduce exact term matches.

\section{Conclusion}
In this work, we present an evaluation framework and a new method for simulated user query variants. Our experiments showed that the retrieval performance of real queries ranges between that of simulated queries from conventional methods based on topic texts and known-items. As a better approximation of user queries, we introduce a simulation method that allows parameterizing the query reformulation behavior and thus better reproduces real queries from specific users. One limitation of our simulations is the exclusion of relevance feedback from previous search results. Users normally include terms of documents or snippets they consider as relevant \cite{DBLP:conf/wsdm/EickhoffTWD14,DBLP:journals/ir/SloanYW15} in their query reformulations. Likewise, the experiments neglect click simulations. We leave it for future work to complement and analyze simulations in this regard.

\appendix
\counterwithin{table}{section}
\counterwithin{figure}{section}
\sectionnotitle{Appendix}
\begin{table}[!b]
  \centering
  \caption{Average retrieval performance over $q$ queries}
  \begin{tabular}{@{}l|g|G|G|G|g|G|G|G|g|G|G|G|@{}}
    \toprule
    \multicolumn{1}{c}{} & \multicolumn{4}{c}{All queries} &  \multicolumn{4}{c}{First queries} &  \multicolumn{4}{c}{Best queries} \\  
    \midrule
    \multicolumn{1}{c}{} &  \multicolumn{1}{c}{$q$} &   \multicolumn{1}{c}{nDCG} &   \multicolumn{1}{c}{P@10} &    \multicolumn{1}{c}{AP} &  \multicolumn{1}{c}{$q$} &   \multicolumn{1}{c}{nDCG} &   \multicolumn{1}{c}{P@10} &    \multicolumn{1}{c}{AP} & \multicolumn{1}{c}{$q$} &   \multicolumn{1}{c}{nDCG} &   \multicolumn{1}{c}{P@10} &    \multicolumn{1}{c}{AP} \\
    \midrule
    UQV\textsubscript{1}       &      150 & .3787 & .4507 & .1581 &       50 & .4293 & .5040 & .2003 &       50 & .4969 & .6320 & .2429 \\
    UQV\textsubscript{2}       &       52 & .4221 & .5058 & .2020 &       50 & .4096 & .4880 & .1894 &       50 & .4103 & .4900 & .1896 \\
    UQV\textsubscript{3}       &       68 & .3922 & .4353 & .1780 &       50 & .3979 & .4560 & .1813 &       50 & .4117 & .4800 & .1878 \\
    UQV\textsubscript{4}       &      123 & .4126 & .4894 & .1888 &       50 & .4469 & .5220 & .2099 &       50 & .5146 & .6300 & .2644 \\
    UQV\textsubscript{5}       &      500 & .3922 & .4330 & .1649 &       50 & .4447 & .4920 & .2043 &       50 & .5353 & .7240 & .2807 \\
    UQV\textsubscript{6}       &      136 & .4030 & .4713 & .1843 &       50 & .4488 & .5080 & .2197 &       50 & .4980 & .5980 & .2515 \\
    UQV\textsubscript{7}       &       50 & .4980 & .5720 & .2418 &       50 & .4980 & .5720 & .2418 &       50 & .4980 & .5720 & .2418 \\
    UQV\textsubscript{8}       &      156 & .3814 & .4545 & .1645 &       50 & .4046 & .4500 & .1799 &       50 & .4556 & .5620 & .2193 \\
    \midrule
    TTS\textsubscript{S1}      &      500 & .0479 & .0306 & .0127 &       50 & .1705 & .1280 & .0541 &       50 & .3066 & .2360 & .0971 \\
    TTS\textsubscript{S2}      &      500 & .1964 & .1716 & .0688 &       50 & .3592 & .3900 & .1604 &       50 & .4391 & .5100 & .2097 \\
    TTS\textsubscript{S$2^\prime$}      &      500 & .3387 & .3426 & .1413 &       50 & .3895 & .4020 & .1821 &       50 & .4639 & .5940 & .2283 \\
    TTS\textsubscript{S3}      &      500 & .3323 & .3632 & .1388 &       50 & .1705 & .1280 & .0541 &       50 & .4776 & .6080 & .2383 \\
    TTS\textsubscript{S$3^\prime$}      &      500 & .3499 & .3874 & .1474 &       50 & .3592 & .3900 & .1604 &       50 & .4709 & .6060 & .2311 \\
    TTS\textsubscript{S4}      &      500 & .4493 & .5168 & .2088 &       50 & .4409 & .4920 & .2072 &       50 & .5945 & .7620 & .3282 \\
    TTS\textsubscript{S$4^\prime$}      &      500 & .4788 & .5626 & .2288 &       50 & .4976 & .5940 & .2429 &       50 & .6207 & .8040 & .3554 \\
    TTS\textsubscript{S$4^{\prime\prime}$}      &      500 & .3780 & .4224 & .1644 &       50 & .4393 & .4860 & .2065 &       50 & .5812 & .7680 & .3222 \\
    \midrule
    KIS\textsubscript{S1}      &      500 & .1334 & .1044 & .0314 &       50 & .2836 & .2040 & .0813 &       50 & .4087 & .4400 & .1492 \\
    KIS\textsubscript{S2}      &      500 & .3969 & .3972 & .1615 &       50 & .5096 & .5400 & .2535 &       50 & .5988 & .7460 & .3429 \\
    KIS\textsubscript{S$2^\prime$}      &      500 & .5114 & .5666 & .2507 &       50 & .5474 & .6220 & .2870 &       50 & .6336 & .7980 & .3762 \\
    KIS\textsubscript{S3}      &      500 & .5598 & .6336 & .3009 &       50 & .2836 & .2040 & .0813 &       50 & .6907 & .8620 & .4299 \\
    KIS\textsubscript{S$3^\prime$}      &      500 & .5941 & .6882 & .3285 &       50 & .5096 & .5400 & .2535 &       50 & .6922 & .8620 & .4337 \\
    KIS\textsubscript{S4}      &      500 & .5216 & .5976 & .2604 &       50 & .5146 & .5960 & .2630 &       50 & .6461 & .8200 & .3902 \\
    KIS\textsubscript{S$4^\prime$}      &      500 & .5008 & .5888 & .2416 &       50 & .5033 & .5980 & .2400 &       50 & .6269 & .8080 & .3703 \\
    KIS\textsubscript{S$4^{\prime\prime}$}      &      500 & .4859 & .5584 & .2293 &       50 & .5191 & .6020 & .2644 &       50 & .6401 & .8360 & .3781 \\
    \bottomrule
  \end{tabular}
  \label{tab:arp}
\end{table}

\begin{figure}[!b]
  \centering
  \begin{subfigure}{.495\textwidth}
    \includegraphics[width=\textwidth]{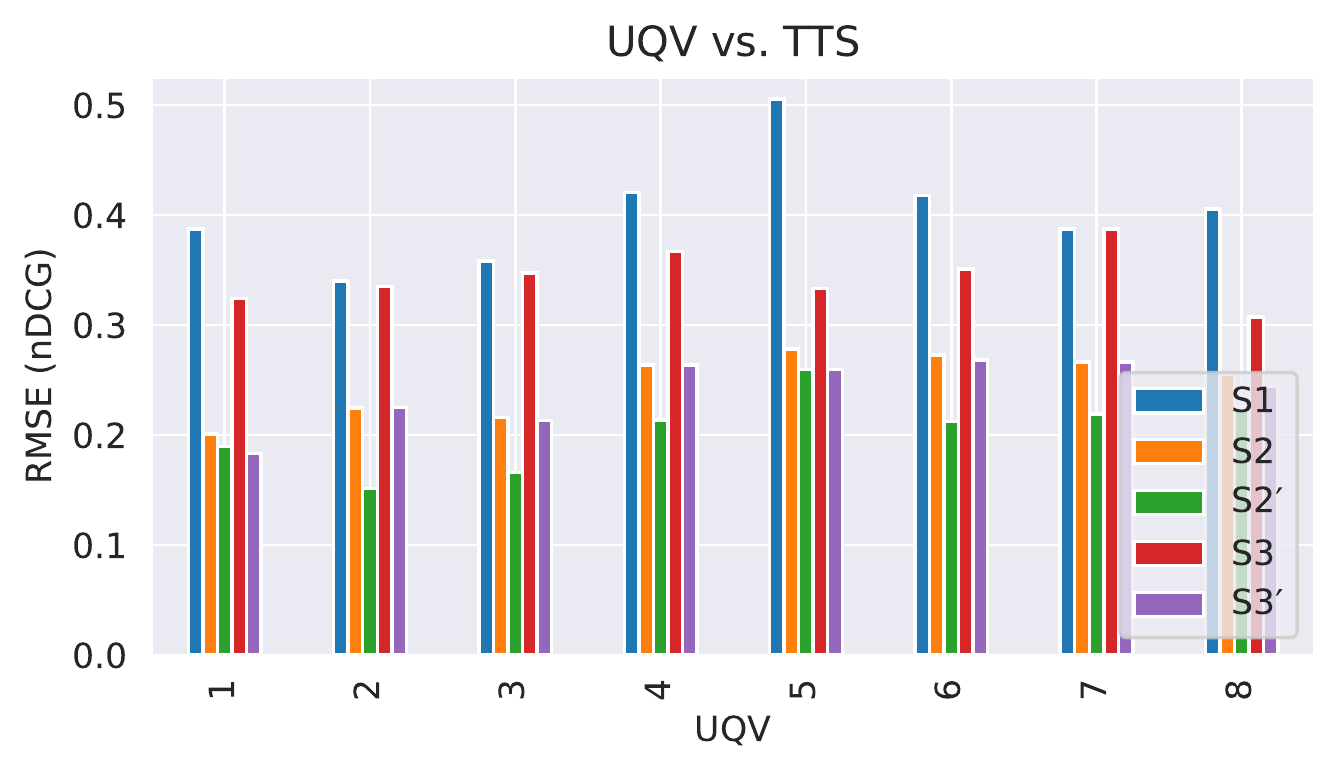}
  \end{subfigure}
  \begin{subfigure}{.495\textwidth}
    \includegraphics[width=\textwidth]{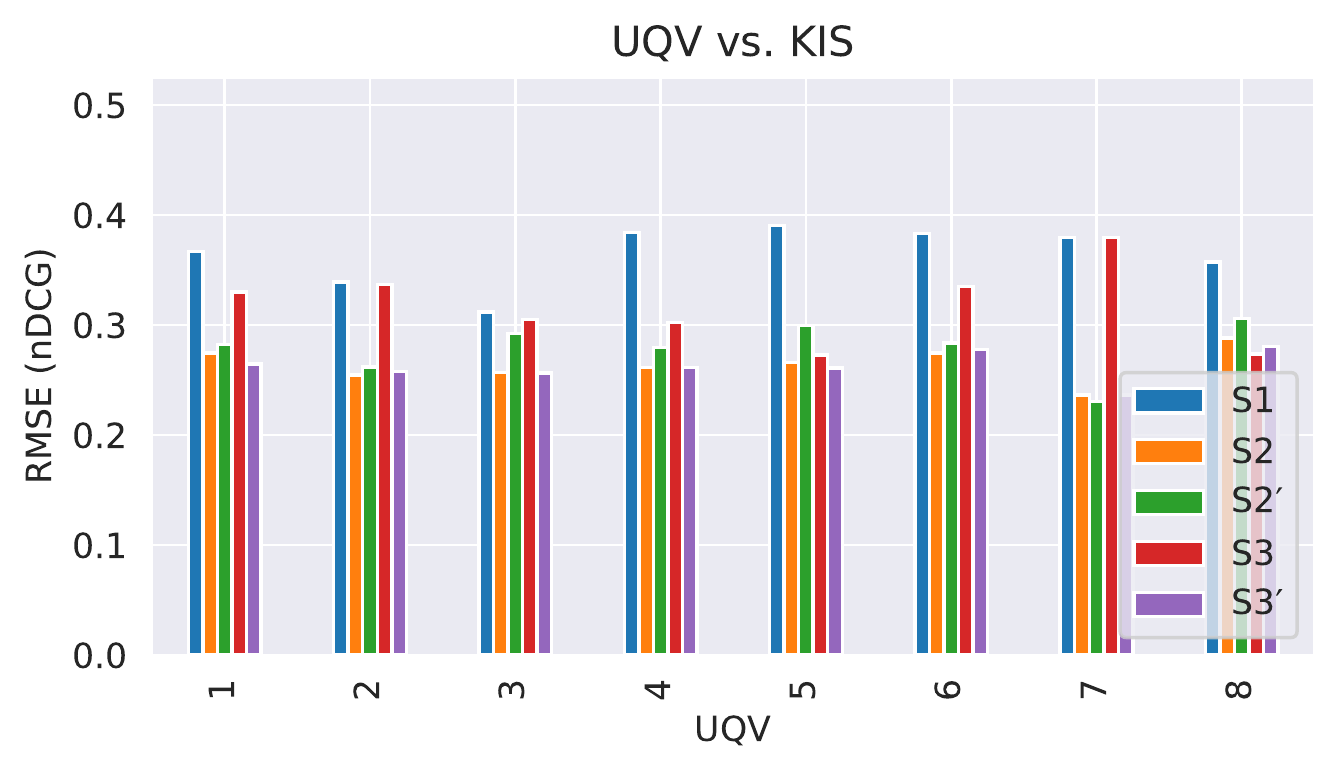}
  \end{subfigure}
  \caption{RMSE between TTS\textsubscript{S1-S$3^\prime$} and KIS\textsubscript{S1-S$3^\prime$} queries and the UQV queries.}
  \label{fig:rmse_s1s5}
  \vspace{5mm}
  \begin{subfigure}{.9\textwidth}
    \includegraphics[width=\textwidth]{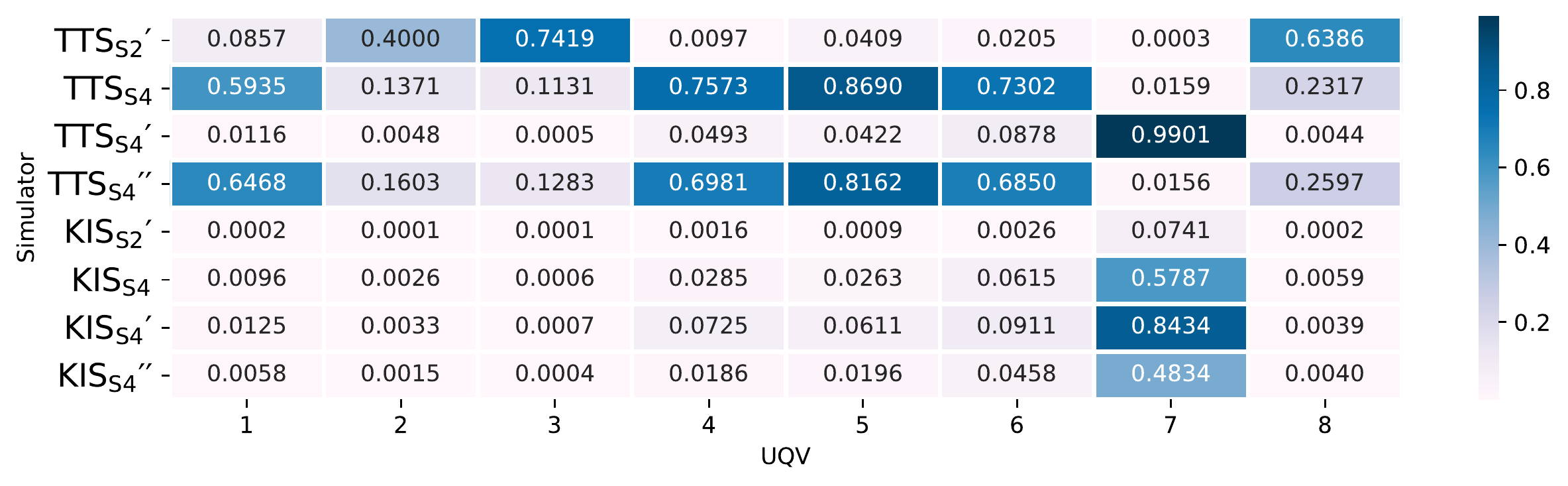}
  \end{subfigure}
  \caption{p-values of paired t-tests between UQV and simulated queries.}
  \label{fig:ttest}
\end{figure}

\bibliographystyle{splncs04}
\bibliography{bibliography-short}

\end{document}